\documentclass[journal]{IEEEtran}
\usepackage{amsmath,graphicx,color,bm}
\usepackage{amssymb}
\usepackage{amsfonts}
\usepackage{amsmath,esint}%
\usepackage{graphicx}
\providecommand{\U}[1]{\protect\rule{.1in}{.1in}}
\newcommand{\Mi}[3]{{\bf M}_{#3} ^{\left(#1\right)} \left( #2 \right)}
\newcommand{\Ni}[3]{{\bf N}_{#3}^{\left(#1\right)} \left( #2 \right)}
\newcommand{\re}[1]{\mbox{Re}\left\{#1\right\}}
\newcommand{\im}[1]{\mbox{Im}\left\{#1\right\}}
\newcommand{\Esp}{{\bf E}_S^{+}}
\newcommand{\Esm}{{\bf E}_S^{-}}
\newcommand{\Ei}{{\bf E}_{i}}

\newcommand{\Mii}[2]{{\bf M}_{#2}^{\left(#1\right)}}
\newcommand{\Nii}[2]{{\bf N}_{#2}^{\left(#1\right)}}

\begin{document}

\title{ Volume Integral Formulation for the Calculation of Material Independent Modes of Dielectric Scatterers }
\author{\IEEEauthorblockN{Carlo Forestiere, Giovanni Miano,
Guglielmo Rubinacci, Antonello Tamburrino, Roberto Tricarico and
Salvatore Ventre}
\IEEEcompsocitemizethanks{\IEEEcompsocthanksitem
C. Forestiere, G. Miano, G. Rubinacci and R. Tricarico are with the
Department of Electrical Engineering and Information Technology, Universit\`{a} degli Studi di Napoli Federico II, via Claudio 21,
 Napoli, 80125, Italy,
 \IEEEcompsocthanksitem A. Tamburrino and S. Ventre are with the Department of Electrical and Information Engineering, Universit\`{a} di Cassino e del Lazio Meridionale, Cassino, Italy
 \IEEEcompsocthanksitem A. Tamburrino is with the Department of Electrical and Computer Engineering, Michigan State University, East Lansing, MI 48824 USA}}

\maketitle	

\begin{abstract}
In the frame of volume integral equation methods, we introduce an alternative representation of the electromagnetic field scattered by a homogeneous  object of arbitrary shape at a given frequency, in terms of a set of modes independent of its permittivity. This is accomplished by introducing an auxiliary eigenvalue problem, based on a volume integral operator. With this modal basis the expansion coefficients of the scattered field are simple rational functions of the permittivity of the scatterer. We show, by studying the electromagnetic scattering from a sphere and a cylinder of  dimensions comparable to the incident wavelength, that only a moderate number of modes is needed to accurately describe the scattered far field.
 
This method can be used to investigate resonant scattering phenomena, including plasmonic and photonic resonances, and to design the permittivity of the object to pursue a  prescribed tailoring of the scattered field. Moreover, the presented modal expansion is computationally advantageous compared to direct solution of the volume integral equation when the scattered field has to be computed for many different values of the dielectric permittivity, given the size and shape of the dielectric body.

\end{abstract}
%
\begin{IEEEkeywords}
Eigenvalues and eigenfunctions, Integral Equations,  Frequency domain analysis, Maxwell Equations, Resonance, Scattering.
\end{IEEEkeywords}
\IEEEpeerreviewmaketitle

\section{Introduction}
The solution of the scattering problem obtained by the majority of analytical and numerical method is often expressed in a form where the contributions of the material and of the geometry are interwoven and cannot be separated. For instance, in the Mie theory for the electromagnetic scattering from spherical homogeneous objects,  the electric size and the permittivity both appear in the argument of the vector spherical wave functions, and, as a consequence, the Mie expansion coefficients are complicated function of their combination. Similar considerations hold true also when the scattering problem is solved by using Volume Integral Equations (VIE) \cite{Schaubert84,draine1994discrete,Sun09,DalNegro09,Markkanen12}, Surface Integral Equations (SIE) \cite{harrington1996field} or the Finite Difference Time Domain (FDTD) \cite{taflove2005computational} solvers. Thus, the complex dependence of the solution of the scattering problem on the permittivity makes the design of the material to achieve assigned constraints on the scattered field complicated. 

Over the years some authors have investigated the possibility of separating the dependence on the material properties from the dependence on the geometry by representing the scattered field in terms of a set of modes independent of the permittivity of the scatter. In the following we denote these modes as Material Independent Modes (MIMs). MIMs have been calculated in the quasi-static limit \cite{Fuchs75,Rojas86,Bergman78,Bergman80,Abajo02,Fredkin2003,Mayergoyz05,DalNegro09}, for the scalar Mie scattering \cite{Markel10}. They have been also derived within the quasi-static \cite{Forestiere09} and retarded \cite{Markel:95} single dipole approximation.

 More recently, MIMs have been derived for the  full-retarded vector scattering by a homogeneous sphere \cite{Forestiere16}, by a coated sphere \cite{pascale2017spectral}, and by a flat slab \cite{Bergman16}. Furthermore, it has been also shown that the investigation of the eigenvalues associated to the MIMs unveils important structural properties of plasmonic and photonic resonances \cite{forestiere2017electromagnetic}. The expansion of the scattered field in terms of MIMs leads to a natural separation of the contributions of the permittivity and of the geometry because the corresponding expansion coefficients are simple rational functions of the permittivity, whereas the MIMs depend solely on the geometry.
This fact has suggested a straightforward methodology to design the permittivity of the object to pursue a  prescribed tailoring of the scattered field, including the cancellation of the backscattering, the suppression of a given multipolar order, and the maximization of the scattered field in the near-field zone \cite{Forestiere16,pascale2017spectral}.

MIMs could also be  useful in solving imaging problems. In particular, it has been already shown that in Eddy Current Tomography the eigenvalues related to the modes are monotonic with respect to the electrical resistivity \cite{Tamburrino15,Tamburrino17}. This property has enabled the use of a real-time imaging algorithm \cite{Tamburrino02}.

Deriving a general and accurate method to compute MIMs of arbitrarily shaped particle is a problem of great importance that has  been addressed so far only in the electrostatic regime \cite{Abajo02,Mayergoyz05,DalNegro09}. In this paper, we tackle this problem by introducing a general approach for the calculation of MIMs in arbitrary shaped homogeneous particles based on a volume integral formulation of the Maxwell's equation.  This approach reduces to the one presented in Ref. \cite{DalNegro09} in the electrostatic limit.
The problem unknown is the sum of the conduction and polarization current densities. It is represented
in terms of its loop and star components. 

The MIMs are not the only modes that can be defined starting  from the full-wave electromagnetic scattering problem in the frequency domain. Different choices are possible resulting in different kinds of modes.  Here, we briefly outline two alternative mode definitions, the quasi normal modes and the characteristic modes, pointing out their profound differences with the MIMs. 

The Quasi Normal Modes (QNMs), also known as  morphology dependent modes and leaky modes  \cite{gastine1967electromagnetic,conwell1984resonant} are often used to investigate open systems. Reviews of QNMs can be found in different contexts, including light scattering \cite{Hill:88} and nanophotonics \cite{kristensen2013}. The QNMs are not orthogonal in the usual sense, moreover they diverge exponentially at large distances \cite{kristensen2013}. Therefore, to be used in any practical application they need to be normalized \cite{kristensen2015normalization}. QNMs of a homogeneous dielectric object do depend on the material, shape and size but are independent of the frequency.
  
Characteristic modes (CMs), also known as  characteristic currents, were first introduced for perfect conductors by the pioneering works of R. J. Garbacz \cite{Garbacz65,garbacz1971generalized}, R. F. Harrington and J. Mautz \cite{harrington1971theory}.
 Later, Harrington et al. have proposed generalization of CMs to structures with dielectric and magnetic materials \cite{harrington1972characteristic,chang1977surface}.
 CMs have been extensively used in the analysis and design of radiation and scattering of two and three dimensional objects \cite{Chen15}, providing guidance for tailoring the excitation of antennas or scatterers to selectively excite desirable radiation modes. CMs have been also used as a set of basis functions for the solution of the scattering problem from an array of small objects \cite{bucci1995use}, resulting in significant reduction of the total number of unknowns compared to a standard MoM approach.  CMs are real and satisfy a weighted orthogonality. Although CMs are independent of the excitation, they do depend on the shape, size, frequency, and material composition of the scattering object.

The layout of the paper is as follows. In Sec. \ref{sec:Model} we derive an auxiliary eigenvalue problem which defines the MIMs starting from a volume integral formulation of the Maxwell's equations.  We also discuss the main properties of the eigenvalues and of the associated modes.  Then, in Sec. \ref{sec:NumericalModel} we introduce the numerical discretization of the unknowns in terms of the star and loop  shape functions. We provide the explicit expression of the matrix equations. Thus, we use the MIMs to expand the scattered field in the presence of an arbitrary external excitation.  In Sec. \ref{sec:Results} we apply the introduced numerical approach to compute the MIMs of a sphere and a cylinder of dimensions comparable to the incident wavelength. In particular, the eigenvalues associated to MIMs of a homogeneous sphere are validated against analogous quantities evaluated through an independent approach \cite{Forestiere16}. The scattering cross section and the radiation diagrams of both sphere and cylinder are obtained starting from MIMs, and compared against the Mie theory and a surface integral formulation of the Maxwell's equations. It is also shown by numerical examples that only few modes are necessary to represent the far field scattered by a dimension comparable to the wavelength. A scenario, in which the presented method is computational advantageous compared to the direct solution of the volume integral equations is also illustrated.

\section{Mathematical Model}
\label{sec:Model}

Let us consider an isotropic and homogeneous material occupying a volume $\Omega$, which is bounded by a closed surface $\partial \Omega$ with outward-pointing normal  $\hat{\bf n}$.
 The medium is non-magnetic with relative permittivity $\varepsilon_R \left( \omega \right) = \varepsilon_R' \left( \omega \right) + i \, \varepsilon_R'' \left( \omega \right)$, surrounded by vacuum. The object is excited by a time harmonic electromagnetic field incoming from infinity $\re{\Ei \left({\bf r}\right) e^{- i \omega t}}$. Let $\Esp$ and $\Esm$ be the scattered electric fields in $\mathring{\Omega}$ and $\mathbb{R}^3 \backslash \bar{\Omega}$, respectively, where $\mathring{\Omega}$ denotes the interior of $\Omega$.

Following Ref. \cite{grat2006}, we now consider the generalized current density in the material region:
\begin{equation}
\mathbf{J}\left(  \mathbf{r}\right)  =
\begin{cases}
   -i\omega \varepsilon_{0} \chi \left( \omega \right)   \left( \Esp \left(  \mathbf{r}\right) +  \Ei \left(  \mathbf{r} \right) \right) &  {\bf r} \in \mathring{\Omega} \\
\mathbf{0} & {\bf r}  \in \mathbb{R}^{3}\backslash\Omega
\end{cases}
\label{eq:CurrentJ}%
\end{equation}
where $\chi \left( \omega \right) = \varepsilon_R  \left( \omega \right) - 1$ is the dielectric susceptibility, and $\varepsilon_{0}$ is the
free space dielectric constant. We choice $\mathbf{J}$ as the problem's unknown. This fact allows us to limit the discretization only to the material regions and to recast the problem as a volume integral equation whose  kernel is the free-space Green function
\cite{ragr1988,ragr1998,grat2006}.
	
We express the electric field produced by $\bf J$ in terms of the vector and scalar potentials, assuming the Lorenz gauge. The vector and scalar potentials can be directly obtained from the current density and the
 surface charge density through the scalar free space Green function $g\left(  \mathbf{r}\right) \triangleq e^{i k_0 r}/\left(  4\pi r\right)$, $k_0= \omega/c$, and $c$ is the speed of light in vacuum. Therefore, the electric field $\Esp$  sustained by $\mathbf{J}$ is:
\begin{multline}
\Esp \left( {\bf r} \right) = +i \omega\mu_{0}\iiint_{\Omega}\mathbf{J}\left(  \mathbf{r}^{\prime
}\right)  g\left(  \mathbf{r-r}^{\prime}\right)  \text{d}V^{\prime} \\ +\frac
{1}{i \omega\varepsilon_{0}}\nabla \varoiint_{\partial \Omega}  \mathbf{J} \left(  \mathbf{r}^{\prime}\right)
\cdot\mathbf{\hat{n}} \left(  \mathbf{r}^{\prime}\right)  g\left(
\mathbf{r-r}^{\prime}\right)  \text{d}S^{\prime} \quad
\forall {\bf r} \in \Omega,
 \label{eq:InducedE}
\end{multline}
where the quantity ${\bf J}$ occurring in the surface integral on the r.h.s. represents the limit of the volume current $\bf J$ along the normal to $\partial \Omega$ as the evaluation point approaches the surface from the internal face of $\partial \Omega$. By combining Eqs. \ref{eq:CurrentJ} and \ref{eq:InducedE}, we obtain the volume integral equation in terms of the unknown $\mathbf{J}$:
\begin{equation}
 \chi^{-1} \, \mathbf{J}\left(  \mathbf{r} \right) - \mathcal{L} \left\{ \mathbf{J} \right\} \left( \mathbf{r} \right) = -i \omega \varepsilon_0 \Ei \left(  \mathbf{r}\right) \quad
\forall {\bf r} \in \Omega,
\label{eq:NonHomogeneous}
\end{equation}
where we have introduced the operator $\mathcal{L} \left\{ \cdot \right\}$
\begin{multline}
\mathcal{L} \left\{ \mathbf{J} \right\} \left( {\bf r} \right) = - \nabla \varoiint_{{\partial \Omega}}  \mathbf{J} \cdot\mathbf{\hat{n}}  \left(  \mathbf{r}^{\prime}\right)  g\left(
\mathbf{r-r}^{\prime}\right)  \text{d}S^{\prime}  \\ + k_0^2 \iiint_{\Omega} \mathbf{J} \left(  \mathbf{r}^{\prime
}\right)  g\left(  \mathbf{r-r}^{\prime}\right)  \text{d}V^{\prime} \qquad \forall {\bf r} \in \Omega.
\label{eq:OperatorL}
\end{multline}
Since the material is assumed to be homogeneous, ${\bf J}$ is divergence free in $\Omega$, whereas its normal component on the domain boundary is proportional to surface charge density. It is straightforward to show that the operator $\mathcal{L}$ is symmetric:
\begin{equation}
   \langle {\bf J}', \mathcal{L} {\bf J}'' \rangle_\Omega = \langle \mathcal{L} {\bf J}',  {\bf J}'' \rangle_\Omega
\end{equation}
where 
 \begin{equation}
\langle \mathbf{A},\mathbf{B} \rangle_V = \iiint_V \mathbf{A} \cdot \mathbf{B} \, \mbox{dV}.
\end{equation}
Aiming at the reduction of the scattering problem to an algebraic form, we introduce the following auxiliary eigenvalue problem
\begin{equation}
  \mathcal{L} \left\{ \mathbf{u} \right\} \left( {\bf r} \right) = \sigma \, \mathbf{u} \left( {\bf r} \right)
  \label{eq:Auxiliary}
\end{equation}
where $\sigma$ is the eigenvalue.
The adjoint of the operator of $\mathcal{L}$ is 
\begin{equation}
\mathcal{L}^\dagger = \mathcal{L}^*
\end{equation}
Therefore, the operator $\mathcal{L}$ is not self-adjoint but symmetric. Its eigenvalues are complex with $\im {\sigma_r^{-1}} < 0$. The eigenmodes ${\bf u}_r$ and ${\bf u}_s$ corresponding to different eigenvalues $\sigma_r$ and $\sigma_s$ are not orthogonal in the usual sense, i.e. $\langle {\bf u}_r^*,{\bf u}_s\rangle_\Omega \ne 0$
 Nevertheless, it can be proved that
\begin{equation}
 \langle {\bf u}_r,{\bf u}_s\rangle_\Omega = 0 \qquad  \text{for}\quad \sigma_{r} \ne \sigma_{s}.
 \label{eq:Orthogonality}
\end{equation}
We denote with $\mathbf{C}_r$ the electric field produced in the whole space by the current ${\bf u}_r$. By considering the volume integral of the quantity $\nabla \cdot \left[ \mathbf{C}_r \times \nabla \times \mathbf{C}_r^* \right]$ over $\mathbb{R}^3 \backslash \bar{\Omega}$,  exploiting the divergence theorem and the properties of the eigenfunctions we readily obtain
\begin{align}
  \re{ \sigma_r^{-1}} &= +\frac{1}{\left\| \mathbf{C}_r \right\|_\Omega^2} \left(  \frac{\left\| \nabla \times \mathbf{C}_r \right\|^2_{\mathbb{R}^3}}{{k_0^2}} -  \left\| \mathbf{C}_r \right\|^2_{\mathbb{R}^3}  \right),   \label{eq:PropertiesEig1}
 \\   \im{ \sigma_r^{-1}} &=   - \frac{1}{\left\| \mathbf{C}_r \right\|_\Omega^2}  \varoiint_{S_\infty} \frac{\left| \mathbf{C}_r \right|^2}{k_0} \, \text{dS},
  \label{eq:PropertiesEig2}
\end{align}
where $\left\| {\bf A} \right\|_{V}^2 = \langle {\bf A}^*, {\bf A} \rangle_{V}$ and $S_\infty$ is an auxiliary closed surface enclosing the scatterer and contained in the far zone.
Equation \ref{eq:PropertiesEig1} suggests that $\re{{ \sigma_r^{-1}}}$ does not have a definite sign, while   Equation \ref{eq:PropertiesEig2} implies that $\im{\sigma_r^{-1}}$ is strictly negative. In particular, $\im{\sigma_r^{-1}}$ is proportional to the contribution of the corresponding mode to the power radiated to infinity, accounting for its radiative losses. 

\section{Numerical Formulation}

\label{sec:NumericalModel}

Following \cite{grat2006}, we split the unknown $\mathbf{J}$ into the sum of its loop and star components, denoted by $\mathbf{J}_{L}$ and $\mathbf{J}_{S}$, respectively, namely:
\begin{equation}
\mathbf{J} = \mathbf{J}_{L} + \mathbf{J}_{S},
\label{eq:CurrentDecomposition}
\end{equation}
where:
\begin{alignat*}{2}
 \nabla \cdot \mathbf{J}_{L} & =0\text{ in } \mathring{\Omega}, \qquad && \mathbf{J}_{L} \cdot\mathbf{\hat{n}}=0\text{ on }\partial\Omega, \\
\nabla\cdot\mathbf{J}_{S}  & =0\text{ in } \mathring{\Omega},  \qquad && \mathbf{J}_{S}%
\cdot\mathbf{\hat{n}}\ne0\text{ on }\partial\Omega.
\end{alignat*}
Thus, we introduce a finite-dimensional approximation of the currents $\mathbf{J}_{L}$, $\mathbf{J}_{S}$ in terms of linear combinations of suitable shape functions, namely $\mathbf{w}_{k}^{L}$'s and $\mathbf{w}_{k}^{S}$'s, respectively. The $k$-th loop shape function $\mathbf{w}_{k}^{L}$ is associated to the $k$-th edge of the finite element discretization of the volume $\Omega$, and it is defined as the curl of the $k$-th edge-element shape functions  $\mathbf{N}_{k}$:
\[
\mathbf{w}_{k}^{L}\left(  \mathbf{r}\right)  =\nabla\times\mathbf{N}%
_{k}\left(  \mathbf{r}\right)
\]

The shape functions $\mathbf{w}_{k}^{S}$'s are used  to discretize the star component $\mathbf{J}_{S}$ describing the effects due to the surface charges appearing on the boundary ${\partial \Omega}$. The function $\mathbf{w}_k^S$ is defined as the curl of the $k$-th edge-element shape function corresponding to an edge on ${\partial \Omega}$.

Then, the unknown current density distribution is represented, at the discrete
level, as
\begin{equation}
\mathbf{J} = \sum_{k=1}^{N_{L}}I_{k}^{L}\mathbf{w}_{k}^{L}+\sum_{k=1}^{N_{S}%
}I_{k}^{S}\mathbf{w}_{k}^{S},
\label{eq:Jdiscretization}%
\end{equation}
where the coefficients $I_{k}^{L}$'s and $I_{k}^{S}$'s 
are the degrees of freedom (DoFs) for the loop and star  components, while $N_{L},\ N_{S}$ are the number of loop and star functions involved in the discretization. Their sum correspond to the total number of degrees of freedom $N_{DoF} = N_L + N_S$.


We obtain the discrete model, by first substituting the current representation of Eq. \ref{eq:Jdiscretization} into the integral equation \ref{eq:NonHomogeneous}, and by then applying the Galerkin method, projecting along the loop and the star shape functions.  Following the outlined steps, we obtain the discretization of the non-homogeneous problem:
\begin{equation}
 \chi^{-1} {\bf R} {\bf I} - {\bf K} {\bf I} = {\bf V}
 \label{eq:NonHomDis}
\end{equation}
where
\begin{equation}
\mathbf{K} = i \omega  \, \mathbf{L} + \frac{1}{i \omega } \mathbf{D},
\label{eq:Zdef}
\end{equation}
and the block matrices and vector are defined as:
\begin{multline*}
\mathbf{R}  = \left[
\begin{array}
[c]{lll}%
\mathbf{R}_{LL} & \mathbf{R}_{LS}  \\
\mathbf{R}_{SL} & \mathbf{R}_{SS}  
\end{array}
\right], \quad
\mathbf{L}  = \left[
\begin{array}
[c]{lll}%
\mathbf{L}_{LL} & \mathbf{L}_{LS} \\
\mathbf{L}_{SL} & \mathbf{L}_{SS} \\
\end{array}
\right], \\
\mathbf{D} = \left[
\begin{array}
[c]{ll}%
\mathbf{0} & \mathbf{0}  \\
\mathbf{0} & \mathbf{D}_{SS} 
\end{array}
\right], 
\quad
\mathbf{I}=\left[
\begin{array}
[c]{c}%
\mathbf{I}^{L}\\
\mathbf{I}^{S}
\end{array}
\right], \quad
\mathbf{V} =\left[
\begin{array}
[c]{c}%
\mathbf{V}^{L}\\
\mathbf{V}^{S}
\end{array}
\right].
\nonumber
\end{multline*}
It is worth to point out that the use of the Galerkin method guarantees that the matrix $\mathbf{K}$, which represents the discretization of the operator $\mathcal{L}$, preserves its symmetry.
The generic occurrences of the block matrices $\bf R$ and $\bf L$ are:
\begin{align*}
\left(  \mathbf{R}_{\alpha\beta}\right)  _{pq}  & = \frac{i}{\omega \varepsilon_0} \iiint_{\Omega}\mathbf{w}%
_{p}^{\alpha} \left(  \mathbf{r}\right) \cdot \mathbf{w}_{q}^{\beta} \left(  \mathbf{r}\right) \text{d}V\\
\left(  \mathbf{L}_{\alpha\beta}\right)  _{pq}  & = \mu_0 \iiint_{\Omega}%
\iiint_{\Omega}\mathbf{w}_{p}^{\alpha}\left(  \mathbf{r}\right)  \cdot
\mathbf{w}_{q}^{\beta}\left(  \mathbf{r}^{\prime}\right)  g\left(
\mathbf{r-r}^{\prime}\right)  \text{d}V\text{d}V^{\prime}
\end{align*}
$\forall \alpha, \beta \in \left\{L, S \right\}$. The non-vanishing elements of the matrix $\mathbf{D}$ are:
\begin{align*}
& \left(  \mathbf{D}_{SS}\right)_{pq}  = \frac{1}{\varepsilon_0} \times \\
& \varoiint_{{\partial \Omega}}\varoiint_{{\partial \Omega}}\left(
\mathbf{w}_{q}^{S}\cdot\mathbf{\hat{n}}^{\prime}\right)  \left(  \mathbf{r}%
^{\prime}\right)  \left(  \mathbf{w}_{p}^{S}\cdot\mathbf{\hat{n}}\right)
\left(  \mathbf{r}\right)  g\left(  \mathbf{r-r}^{\prime}\right)
\text{d}S^{\prime}\text{d}S.
\end{align*}
The occurrences of the vector $\mathbf{V}$ are:\begin{equation*}
\left(  \mathbf{V}^{\alpha}\right)  _{p}  \mathbf{=}\iiint_{\Omega}%
\mathbf{w}_{p}^{\alpha}\cdot\mathbf{E}_{i}\text{d}V, \quad \forall\alpha  \in \left\{L,S \right\}.
\end{equation*}

In absence of external excitation from Eq. \ref{eq:NonHomDis} we obtain the discrete eigenvalue problem
\begin{equation}
{\bf K} {\bf I} = \sigma {\bf R} {\bf I},
\label{eq:LinearEigProblem}
\end{equation}%
where $\sigma$ is the eigenvalue,
We denote the  numerical eigenvalues with $\sigma_h$ and the corresponding eigenvectors with ${\bf I}_h$. The eigenvectors are not orthogonal in the usual sense but we have:
\begin{equation}
 {\bf I}_h^T {\bf R} {\bf I}_k = 0 \qquad \sigma_h \ne \sigma_k.
\end{equation}
By expanding the solution ${\bf I}$ of the discrete problem of Eq. \ref{eq:NonHomDis} in terms of the eigenvalues $\sigma_h$ and the eigenvectors ${\bf I}_h$ of Eq. \ref{eq:LinearEigProblem} we obtain:
\begin{equation}
   {\bf I} =  \sum_{h=1}^{N_{DoF}} \frac{1}{\chi^{-1} - \sigma_h } \;
   \frac{{\bf I}_h^T {\bf V}}{  {\bf I}_h^T {\bf R} {\bf I}_h} \; {\bf I}_h
   \label{eq:ModeExpansion}
\end{equation}
In Eq. \ref{eq:ModeExpansion} the dependence of ${\bf I}$ from the material and the geometry are naturally separated.
The eigenvalues $\sigma_h$ and the eigenvectors ${\bf I}_h$ are  permittivity independent, and they depend on the shape and size of the dielectric object, and on the frequency. The susceptibility appears in the multiplicative factors only as $ 1/\left( \chi^{-1} - \sigma_h \right)$.

For passive materials with non-negative imaginary part of the susceptibility we have $\im{\chi} \ge 0$, thus the quantity  $\left| \chi^{-1} - \sigma_h   \right|$ in Eq. \ref{eq:ModeExpansion} does not vanish because $\im{\sigma_{h}^{-1}}<0$. Nevertheless, the amplitude of the h-th mode increases as the distance between $\chi^{-1}$ and $\sigma_h$ is reduced. In other words, Eq. \ref{eq:ModeExpansion} exemplifies that, for a fixed frequency, when the scatterer's material closely ``matches'' an eigenvalue, the corresponding mode undergoes a boosting, namely a ``resonance'' in a ``material picture''. This picture is dual with respect to the usual ``frequency picture'', where the material is instead fixed and the frequency plays the role of spectral parameter.
The ``material picture'' is particularly relevant in light of the latest advances in Metamaterials' design and fabrication techniques, which are enabling the effective value of material's permittivity and permeability to be engineered with increasing precision.

 Moreover, when $\re{\sigma_{h}^{-1}}<-1$ the amplitude of the h-th mode will be particularly strong for materials with negative permittivity, in this case we denote the corresponding mode as a ``plasmonic'' mode. On the contrary when $\re{\sigma_{h}^{-1}}>-1$ will be particularly strong for materials with positive permittivity, in this case we denote the corresponding mode as ``photonic''.

In the next sections, we refer to the introduced method as MIM-VIE.

\section{Numerical Results}
\label{sec:Results}


We carried out the computation presented in this section on a cluster of 25 Intel Xeon CPU E5-2690 cores operating at 2.90GHz and equipped with 128 GB of RAM memory.
The Fortran library ARPACK \cite{lehoucq1998arpack}  has been used for the calculation of the generalized eigenvalues and eigenvectors of Eq. \ref{eq:LinearEigProblem}. These routines are based upon an algorithmic variant of the Arnoldi process called the Implicitly Restarted Arnoldi Method (IRAM) \cite{lehoucq1998arpack}. In particular, we compute   $N_{SR}$ eigenvalues of smallest real part, $N_{LI}$ eigenvalues of largest imaginary part, $N_{SI}$ eigenvalues of smallest imaginary part, and we consider the union of these three sets, whose dimension is denoted with $N_{tot}$. We set $N_{SR}=N_{LI}=N_{SI}=200$ and a converge tolerance $\tau = 10^{-2}$.  It is apparent that $N_{tot}  \le N_{SR} + N_{LI} + N_{SI}$, because these three sets may have finite intersection. It is worth to point out that the set of eigenvalues with the largest real part have not been computed since they are associated to modes of very high order playing no role in the scattering process.
  
Subsequently, we filter the modes by retaining only those associated to the eigenvalues located outside a box centred in the origin of the complex plane, satisfying simultaneously the following criteria
\begin{equation}
\begin{aligned}
  \left| \re{\sigma_h} \right| > \xi \cdot \, \max_{k=1\ldots N_{tot}}{\left| \re{\sigma_k} \right|}   \\
  \left| \im{\sigma_h} \right| > \xi \cdot \, \max_{k=1\ldots N_{tot}}{\left| \im{\sigma_k} \right|}  
\end{aligned}
\label{eq:Filter}
\end{equation}
where we assumed different values of $\xi$. We denote with $h_{max}$ the total number of modes which pass this filter. In this section, we restrict the sum of Eq. \ref{eq:ModeExpansion} to only these $h_{max}$ MIMs.

In this section, we investigate the scattering cross section and the radiation patterns of a sphere and a cylinder by using the MIMs expansion. They are excited by a plane wave of unit intensity polarized along $\hat{x}$ and propagating along $\mathbf{\hat{z}}$,
\begin{equation}
  {\bf E}_i =   e^{ i k_0 z} \, \hat{\bf x}
  \label{eq:PW}
\end{equation}
In particular, the scattering cross section is \cite{bohren08,WriedtBook}:
\begin{equation}
   C_{sca} =  \frac{c_0}{\omega} \varoiint_{S_c} \hat{\bf e}_r \cdot \mbox{Im} \left\{    \left(  \boldsymbol{\nabla} \times \Esm \right)^* \times \Esm  \right\} \, \mbox{dS},
   \label{eq:ScattCrossSec}
\end{equation}
where $S_c$ is an auxiliary surface enclosing $\Omega$. 
It is worth noting that interference among several MIMs may take place  in the total scattered power, because the MIMs are not orthogonal\cite{forestiere2017electromagnetic}.
The radiation pattern is  \cite{WriedtBook}
\begin{equation}
 \mathbf{E}_S^\infty \left( \theta, \phi \right) = \displaystyle\lim_{r \rightarrow \infty}  \left[ r{e^{- i k_0 r}} \Esm \left( { r},\theta,\phi  \right)\right].
 \label{eq:RadiationPattern}
\end{equation}

\subsection{Sphere}
\begin{figure}[!t]
\centering
\includegraphics[width=\columnwidth]{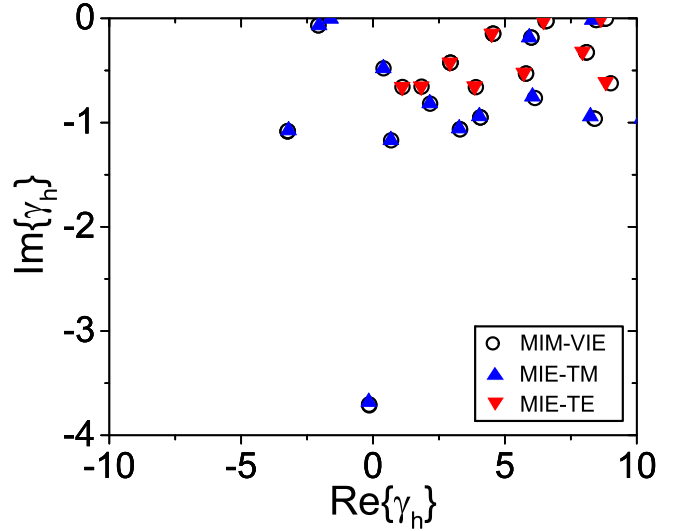}
\caption{Quantities $\gamma_h = \sigma_h^{-1}+1$ represented in the complex plane, where $\sigma_h$ are the eigenvalues of Eq. \ref{eq:LinearEigProblem} (black open circle) for a sphere with diameter $D=\lambda$. Analogous quantity evaluated with the approach of Ref. \cite{Forestiere16} for the TM (blue up-pointing triangles) and TE (red down-pointing triangles) modes.} 
\label{fig:EigSphere}
\end{figure}

Let us consider a homogeneous sphere of diameter $D=\lambda$, where $\lambda$ is the wavelength in vacuum.
We considered an hexahedral volume mesh with $N_p=26075$ points and $N_e=24800$ elements, leading to $N_L=48401$ loop shape functions, and $N_S=2399$ star shape functions, corresponding to a total of $50800$ complex unknowns.

First, we validate the calculation of the eigenvalues $\sigma_h$ against the corresponding quantities calculated by using the approach introduced in Ref. \cite{Forestiere16}. The latter approach consists in numerically finding the the roots of a polynomial, whose coefficients are analytically known  for a sphere. For simplicity of representation, we compare the quantities $\gamma_h = \sigma_h^{-1}+1$. 
In Fig. \ref{fig:EigSphere}, we show with a black circle the $\gamma_h$'s associated to the eigenvalues calculated with the approach introduced in this paper, with a red down-pointing triangle the same quantity associated to magnetic (TE) modes, and with a blue up-pointing triangle the eigenvalues associated to electric (TM) modes, both calculated by using Ref. \cite{Forestiere16}. Only the $\gamma_h$'s belonging to the box of the complex plane $\left[-10, 10\right] \times \left[-4, 0 \right]$ are shown. They are obtained by using the filter of Eq. \ref{eq:Filter} with $\xi=10^{-3}$ are shown.  A small error is appreciable only for moderately positive $\text{Re} \left\{ \gamma_h \right\}$, where the MIM-VIE method tends to  overestimate the real part of $\gamma_h$. This error does not improve by increasing the tolerance $\tau$ up to $10^{16}$, while the eigenvalues' positions remain the same.  It is worth noting that the quantities shown in Fig. \ref{fig:EigSphere} depend neither on the excitation conditions nor on the permittivity $\varepsilon_r$, but they solely depend on the quantity $D/\lambda$.

\begin{figure}[!t]
\centering
\includegraphics[width=\columnwidth]{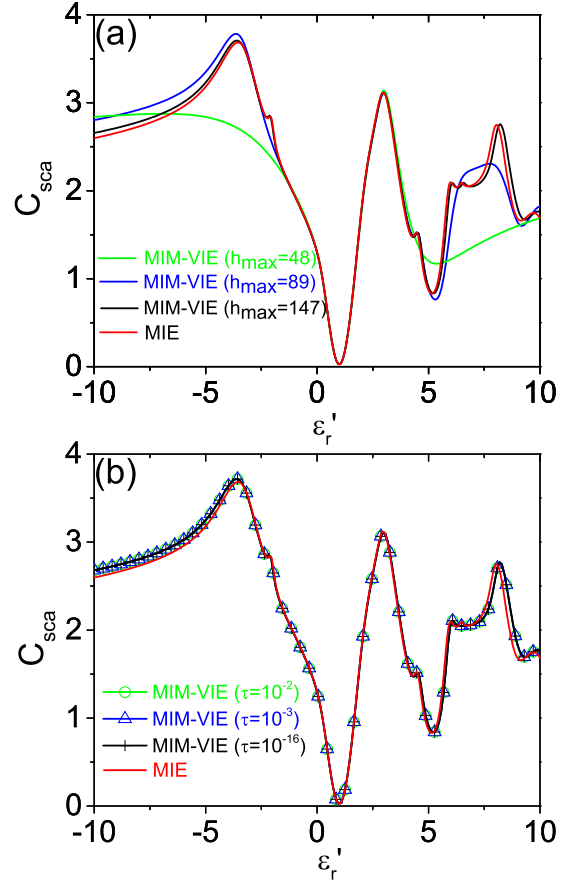}
\caption{Scattering Cross Section $C_{sca}$ of an isolated sphere of diameter $D=\lambda=1m$ with permittivity $\varepsilon_r=\varepsilon_r'+ i \, \varepsilon_r''$ excited by a linearly polarized plane wave as a function of $\varepsilon_r'$. We assumed $\varepsilon_r''=0.1$. In panel (a) we computed $C_{sca}$ using the VIE approach with $h_{max}=48$ (green line), $h_{max}=89$ (blue line), and $h_{max}=147$ (black line) MIMs, obtained using the filter of Eq. \ref{eq:Filter} with $\xi=5 \cdot 10^{-2}$, $\xi=5 \cdot 10^{-3}$, and $\xi= 10^{-3}$, respectively, and assuming $\tau=1\cdot 10^{-2}$, and by the standard Mie theory by using $n_{max}=10$ (red line).
In panel (b) we computed $C_{sca}$ using the VIE approach with $h_{max}=147$ using several tolerances, namely $\tau=  10^{-2}$, $\tau=10^{-3}$,  $\tau= 10^{-16}$, and by the standard Mie theory by using $n_{max}=10$ (red line).
}
\label{fig:CscaSphere}
\end{figure}

Next, in Fig. \ref{fig:CscaSphere} we show the scattering cross section $C_{sca}$ of the sphere as a function of the real part of its permittivity $\varepsilon_r' \in \left[ - 10, 10 \right]$, whereas the imaginary part is fixed to the value $\varepsilon_r''=0.1$. The sphere is excited by the plane wave of Eq. \ref{eq:PW} with $\lambda=1m$. We computed $C_{sca}$ using Eqs. \ref{eq:ScattCrossSec} and \ref{eq:ModeExpansion}. As a reference solution we use the standard Mie Theory \cite{Bohren1998}, where the scattered electric field has been expanded in terms of vector spherical wave functions (VSWFs):
\begin{equation}
   \Esm = \sum_{n=1}^{n_{max}} E_n \left( i a_n \Ni{3}{k_0 {\bf r}}{e1n} - b_n \Mi{3}{k_0 {\bf r}}{o1n} \right)
\end{equation}
where $a_n$ and $b_n$ are the Mie scattering coefficients, which can be found in Ref. $\cite{Bohren1998}$ and $E_n = i^n(2n +1)/\left[n(n +1)\right]$ and the cross section is obtained by
\begin{equation}
  C_{sca} = \frac{2\pi}{k_0^2} \sum_{n=1}^{n_{max}} \left( 2n + 1\right) \left( \left| a_n \right|^2 + \left| b_n \right|^2  \right)  
\end{equation}
where we assumed $n_{max}=10$. The functions $\Nii{3}{ emn}$  and $\Mii{3}{ omn}$
are the VSWFs of the radiative kind and the subscripts $e$ and $o$ denote even and odd azimuthal dependence. 

Specifically, in Fig. \ref{fig:CscaSphere} (a) we investigate the convergence of $C_{sca}$ as we increase the number of considered MIMs. In particular, we performed the filtering of the modes by using Eq. \ref{eq:Filter}, and assuming $\xi= 5 \cdot 10^{-2}$, $\xi= 5 \cdot 10^{-3}$, and $\xi= 10^{-3}$, thus retaining $h_{max}=48$ (green line),  $h_{max}=89$ (blue line), and $h_{max}=147$ (black line) eigenvalues, respectively.  
  We assumed a tolerance of $\tau= 10^{-2}$. We compare the solutions with the standard Mie theory by using $n_{max}=10$ (red line).
It is apparent that $48$ modes are able to satisfactory describe the $C_{sca}$  only within a small interval centred at $\varepsilon_r'=1$. Then, by considering $h_{max}=89$ MIMs we note an increase of the permittivities'interval in which the VIE solution agrees well with the MIE solution. Eventually, for $h_{max}=147$ we achieved a very good agreement within the whole investigated interval. We also appreciate a slight shift of the peaks of the MIM-VIE solution with respect to the reference Mie solution for moderate positive permittivities, due to the overestimation of the real part of $\gamma_h$, already apparent in Fig. \ref{fig:EigSphere}.  

In Fig. \ref{fig:CscaSphere} (b) we investigate the accuracy of the computed $C_{sca}$ as a function of the tolerance $\tau$. In particular, we considered $\tau= 10^{-2}$, $\tau=  10^{-3}$, and $\tau=  10^{-16}$. The number of modes was instead fixed to $h_{max}=147$ ($\xi= 10^{-3}$). We compare the solutions with the standard Mie theory by using $n_{max}=10$ (red line). It is apparent that with tolerance of $\tau= 10^{-2}$ we achieve a good agreement with the Mie theory. A further increase of the tolerance does not appreciably improves the solution.
\begin{figure}[!t]
\centering
\includegraphics[width=\columnwidth]{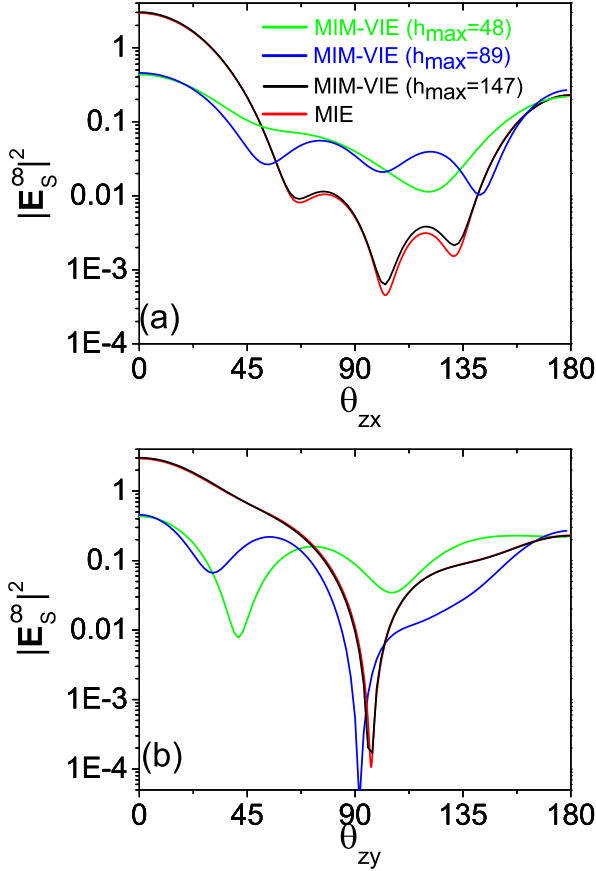}
\caption{Radiation diagram of the electric field scattered from the homogeneous sphere with diameter $D = \lambda = 1m$, as a function of the inclination angle $\theta$ in $zx$ plane (a) and in the $yz$ plane (b). We computed the solution exploiting the VIE approach with $h_{max}=48$ (green line), $h_{max}=89$ (blue line), and $h_{max}=147$ (black line) MIMs, obtained using the filter of Eq. \ref{eq:Filter} with $\xi=5 \cdot 10^{-2}$, $\xi=5 \cdot 10^{-3}$, and $\xi= 10^{-3}$, respectively, and by the standard Mie theory by using $n_{max}=10$ (red line).}
\label{fig:dSCS_Sphere}
\end{figure}

Next,  we show in Fig. \ref{fig:dSCS_Sphere} the  radiation pattern intensity $ \left| \mathbf{E}_S^\infty \left( \theta, \phi \right) \right|^2$ as a function of the inclination angle $\theta$ in the $zx$  plane (a), (c) and in the $yz$  plane (b), (d). In particular, we investigate the convergence of $ \left| \mathbf{E}_S^\infty \left( \theta, \phi \right) \right|^2$ as a function of the number of considered MIMs.
 The sphere is excited by the plane wave of Eq. \ref{eq:PW} with $\lambda=1$. The angles $\theta=0$ and $\theta=\pi$ correspond to the directions of forward- and back- scattering, respectively. We assumed a permittivity $\varepsilon_r = 5 + 0.1 i$, and we performed the calculation using Eq. \ref{eq:ModeExpansion} with $h_{max}=48$ (green line), $h_{max}=89$ (blue line), and $h_{max}=147$ (black line) MIMs and by the standard Mie theory (red line). We obtained good agreement only for $h_{max}=147$.

In conclusion, the introduced approach correctly evaluates the eigenvalues $\sigma_h$ of a sphere of diameter $D=\lambda$. Moreover, it is also capable to accurately compute the scattering cross section and the radiation diagrams by using only a moderate number of MIMs.

We now show that the MIM-VIE solution may be computational advantageous in a scenario relevant to practical applications if compared to a direct solution of the VIE formulation of Eq. \ref{eq:NonHomDis}, namely
\begin{equation}
  {\bf I} = \left( \chi^{-1} {\bf R} - {\bf K} \right)^{-1} {\bf V}.
\end{equation}
First, let us derive the time needed for the calculation of the scattering response at a given frequency of a number $n_{\varepsilon_r}$ of different spheres of given size and shape, but having  $n_{\varepsilon_r}$ different values of $\varepsilon_r$, by using both the MIM-VIE and the direct VIE solvers. The cumulative time for assembling both matrices $\bf K$ and $\bf R$ on $25$ processors is the same for both approaches, namely $ t_{\text{ASM}} = 2h \, 12m $. The total time needed for the calculation of the sets of eigenvalues and corresponding MIMs is $t_{\text{EIG}} = 44 m $. The total time needed for the MIM-VIE formulation is with good approximation independent of $n_{\varepsilon_r}$, being approximatively $t_{\text{MIM-VIE}} \approx t_{\text{ASM}} + t_{\text{EIG}}$.
On the contrary, using the direct VIE solver, we have to find the solution of a different linear system for each of the $n_{\varepsilon_r}$ values of permittivity. We accomplished this task by LU factorization using the ScaLAPACK routine called PZGETRF  \cite{blackford1997scalapack}. Each inversion requires a time $t_{\text{FCT}} = 11m$.
Therefore, the total time is $t_\text{VIE} = t_{\text{ASM}} + n_{\varepsilon_r} \times t_{\text{FCT}}$. The { break-even} value of $n_{\varepsilon_r}$, in correspondence of which the MIM-VIE method becomes computationally advantageous compared to a direct VIE solution, is  $n_{\varepsilon_r} \approx 4$. It is also worth to point out that if we are interested in finding the scattering response of the object for just one value of permittivity, i.e. $n_{\varepsilon_r} = 1$, the MIM-VIE solution is roughly $1.23$ times slower than the direct VIE solution. This analysis clearly demonstrates that the
VIE-MIM approach is particularly suited for optimizing the the material permittivity.

\subsection{Cylinder}
\begin{figure}[!t]
\centering
\includegraphics[width=\columnwidth]{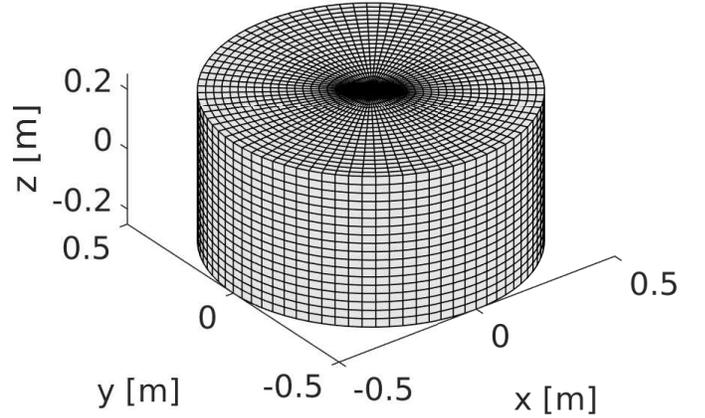}
\caption{Hexahedral volume mesh of the investigated cylinder (diameter $D=\lambda$, height $H=\lambda/2$). It features $37183$ points and $34488$ elements, which correspond to $N_L=66341$ loop shape functions, and $N_S=5271$ star shape functions, corresponding to a total of $71612$ complex unknowns. }
\label{fig:CylMesh}
\end{figure}

We now consider a cylinder of diameter $\lambda$ and height $H=0.5 \lambda$.
The finite elements mesh (hexahedral elements) is shown in Fig. \ref{fig:CylMesh}, whose details are reported in the figure caption.
\begin{figure}[!t]
\centering
\includegraphics[width=\columnwidth]{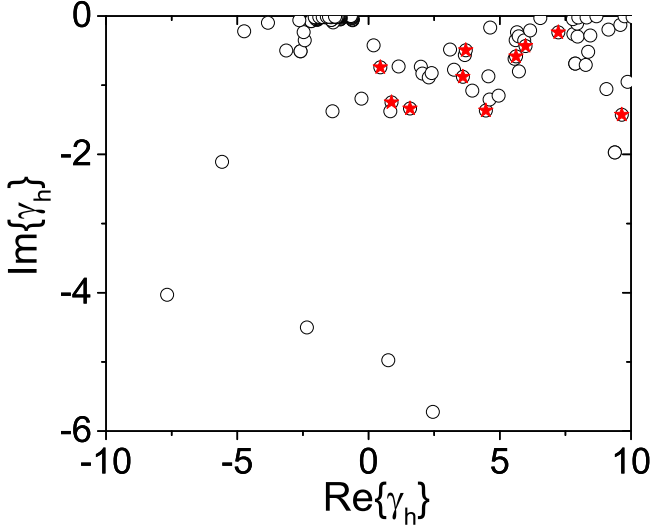}
\caption{Quantities $\gamma_h = \sigma_h^{-1}+1$ represented in the complex plane, where $\sigma_h$ are the eigenvalues of Eq. \ref{eq:LinearEigProblem} (black open circle) for cylinder with basis diameter $D=\lambda$ and height $H=0.5\lambda$. With red stars we also represent the $\gamma_h$ associated to the modes shown in Fig. \ref{fig:Modes_Cyl} and listed in Tab. \ref{tab:Cyl}.}
\label{fig:Eig_Cyl}
\end{figure}

\begin{table*}
\centering
\caption{Values of $\gamma_h = \sigma_h^{-1}+1$ associated to the eigenvalues $\sigma_h$ of the ten modes shown in Fig. \ref{fig:Modes_Cyl}.}
\begin{tabular}{|c||c|c|c|c|c|c|c|c|c|c}
\hline
\# & 1 & 2 & 3 & 4 & 5 \\    
\hline
$\gamma_h$ & $ 0.45 - 0.74 i $ & $ 0.88 - 1.2 i $ & $ 1.6 - 1.3 i $ & $ 3.6 - 0.88 i $ & $ 3.7 - 0.49 i $ \\
\hline
\hline
\# & 6 & 7 & 8 & 9 & 10 \\    
\hline
$\gamma_h$ &  $ 4.5 - 1.4 i $ & $ 5.6 - 0.58 i $ & $ 6.0 - 0.43 i $ & $ 7.2 - 0.24 i $ & $ 9.6 - 1.4 i $ \\
\hline
\end{tabular}
\label{tab:Cyl}
\end{table*}

In Fig. \ref{fig:Eig_Cyl} we show  with black circles the quantity $\gamma_h = \sigma_h^{-1}+1$. We used a tolerance $\tau=10^{-2}$, and by using a filter with $\xi = 10^{-3}$ we only retain $h_{max}=208$ eigenvalues. Only the $\gamma_h$'s belonging to the box of the complex plane $\left[-10, 10\right] \times \left[-6, 0 \right]$ are shown. 

\begin{figure}[!t]
\centering
\includegraphics[width=\columnwidth]{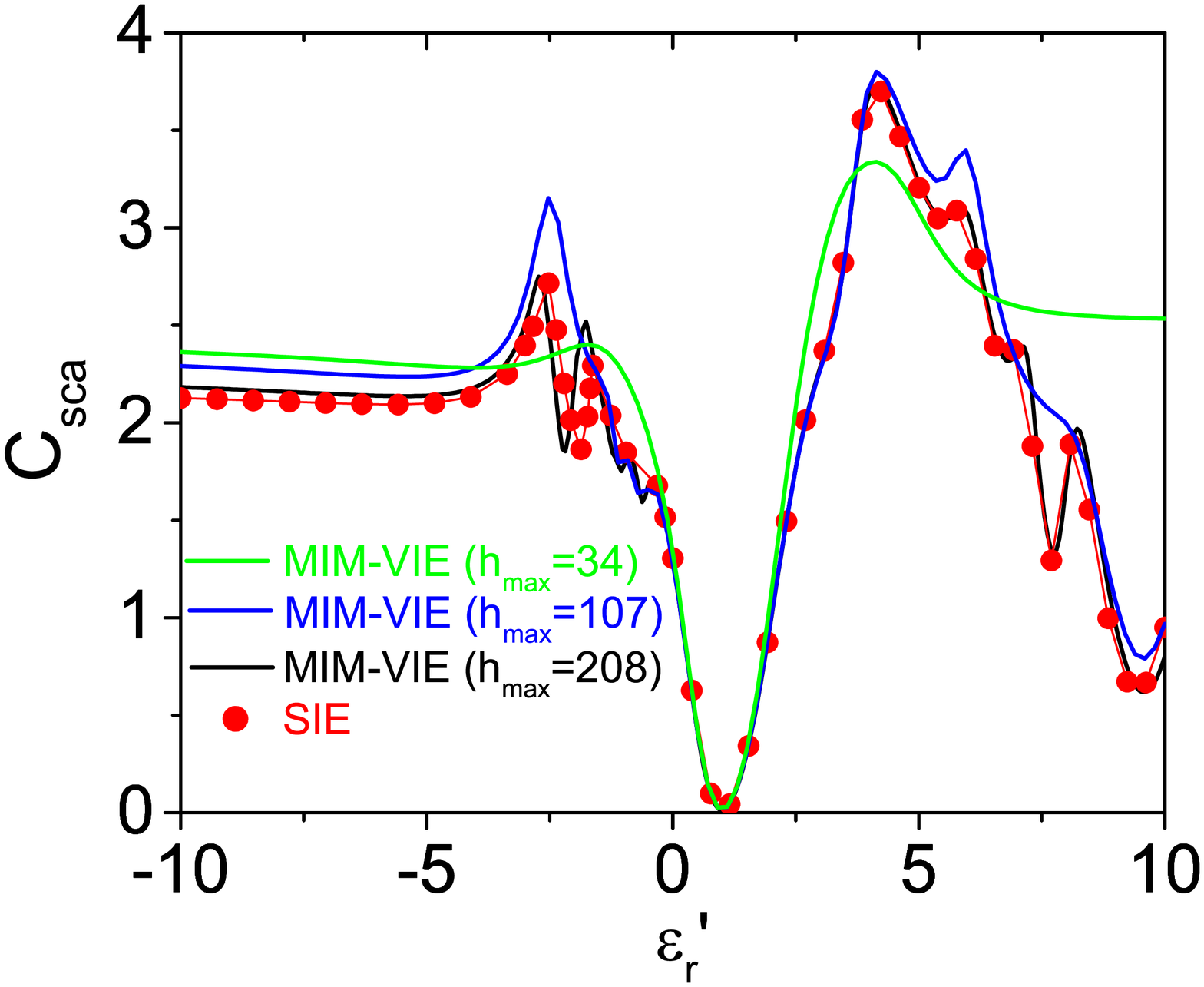}
\caption{Scattering Cross Section $C_{sca}$ of an isolated cylinder of diameter $D=\lambda$ and height $H=\lambda/2$ with permittivity $\varepsilon_r=\varepsilon_r'+ i \, \varepsilon_r''$ excited by a linearly polarized plane wave as a function of $\varepsilon_r'$. We assumed $\varepsilon_r''=0.1$. We computed $C_{sca}$ using the VIE approach with $h_{max}=34$ (green line), $h_{max}=107$ (blue line), and $h_{max}=208$ (black line) MIMs, obtained using the filter of Eq. \ref{eq:Filter} with $\xi=5 \cdot 10^{-2}$, $\xi=5 \cdot 10^{-3}$, and $\xi= 10^{-3}$, respectively, and by using the PMCHWT surface integral formulation (red line).}
\label{fig:Csca_Cyl}
\end{figure}

\begin{figure*}[!t]	
\centering
\includegraphics[width=\textwidth]{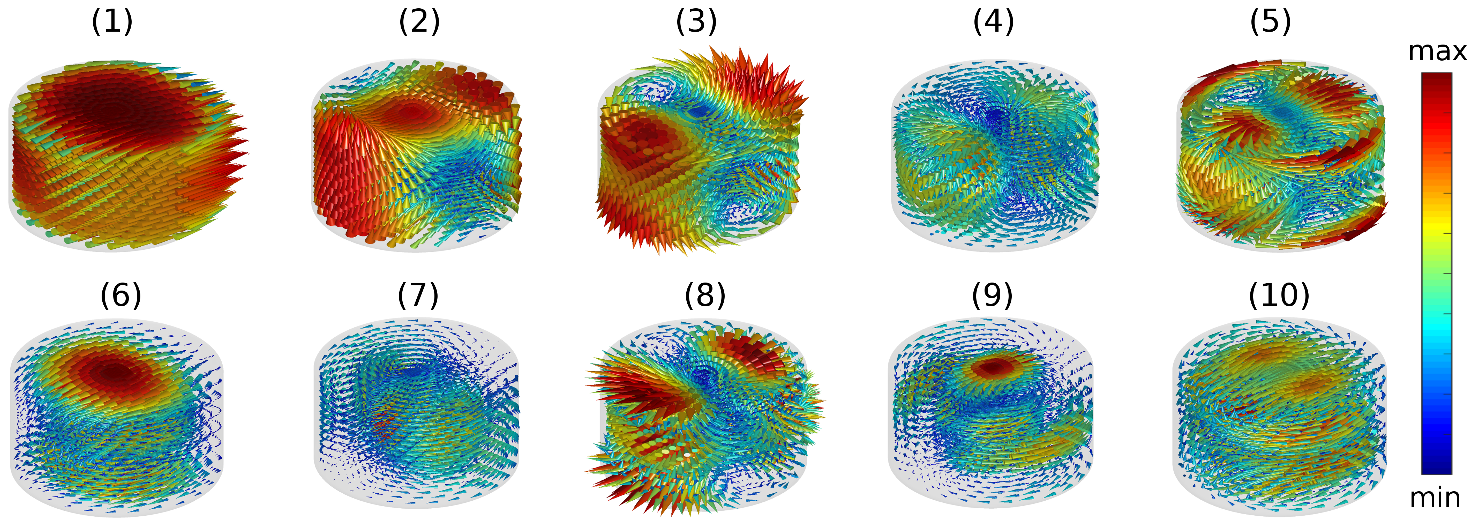}
\caption{Cone plot of the real part of the modes that dominate the total scattered power of a cylinder with $D=\lambda$, $H=0.5\lambda$ when $\varepsilon_r=4.135 + 0.1 j$.}
\label{fig:Modes_Cyl}
\end{figure*}

Next, we investigate the convergence of the scattering cross section $C_{sca}$ by varying the number of considered MIMs. In particular, in Fig. \ref{fig:Csca_Cyl} we plot the  $C_{sca}$ of the investigated homogeneous cylinder as a function of the real part of its permittivity $\varepsilon_r' \in \left[ - 10, 10 \right]$, whereas the imaginary part is fixed to the value $\varepsilon_r''=0.1$. The cylinder is excited by the plane wave of Eq. \ref{eq:PW} with $\lambda=1m$.  As reference solution we use the solution of the  Poggio-Miller-Chang-Harrington-Wu-Tsai (PMCHWT) Surface Integral Formulation of the Maxwell's equations \cite{harrington1996field} by RWG basis functions using a triangular mesh with  $N_{p} = 1373$ nodes, $N_t = 2742$ triangular elements,  $N_{e} = 4113$ edges, corresponding to $N_{DoF}=8226$ degrees of freedom.
First, we computed $C_{sca}$ using the VIE approach with $h_{max}=34$ (green line). As for the case of a sphere, we notice a good agreement only within a small interval centred at $\varepsilon_r'=1$. The convergence improves when $h_{max}=107$ modes are considered. Finally, by using $h_{max}=208$ MIMs we can appreciate a very good agreement. Only for negative values of $\varepsilon_r'$, the two plasmonic resonances appear to be slightly shifted with respect to the surface integral formulation. We plot in Fig. \ref{fig:Modes_Cyl} the ten modes that play the most relevant role in correspondence of the peak at $\varepsilon_r = 4.135 + 0.1 j$. The values of $\gamma_h$ associated to these eigenvalues are tabulated in Tab. \ref{tab:Cyl}, and also plotted with red stars in Fig. \ref{fig:Eig_Cyl}. By considering exclusively these 10 modes in Eq. \ref{eq:ModeExpansion}, we would only commit an error of $9\%$ in the estimation of $C_{sca}$.

\begin{figure}[!t]
\centering
\includegraphics[width=\columnwidth]{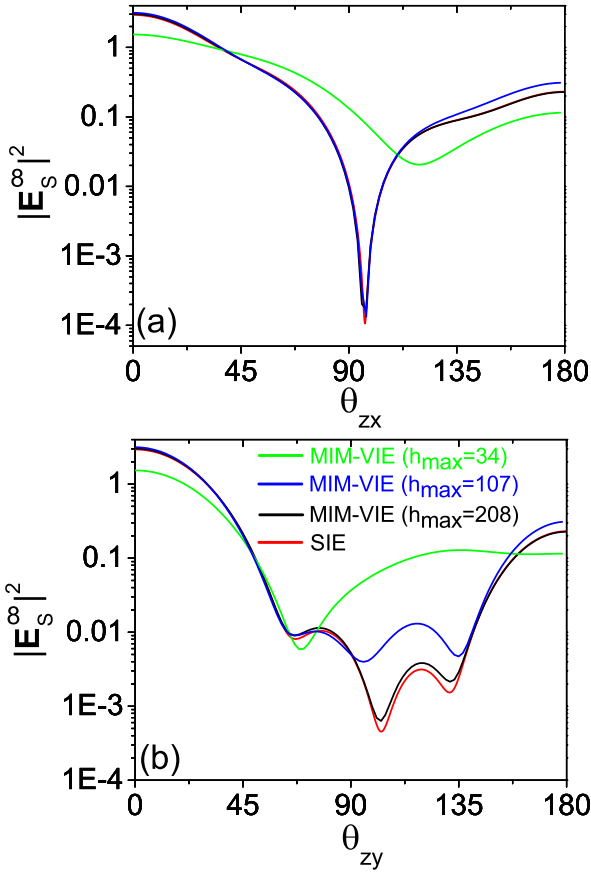}
\caption{Radiation diagram of the electric field scattered from the homogeneous cylinder with basis diameter $D = \lambda = 1m$ and height $H = \lambda/2=0.5m$, as a function of the inclination angle $\theta$ in $zx$ plane (a) and in the $yz$ plane (b). We computed $C_{sca}$ using the VIE approach with $h_{max}=34$ (green line), $h_{max}=107$ (blue line), and $h_{max}=208$ (black line) MIMs, obtained using the filter of Eq. \ref{eq:Filter} with $\xi=5 \cdot 10^{-2}$, $\xi=5 \cdot 10^{-3}$, and $\xi=10^{-3}$, respectively, and by using the PMCHWT surface integral formulation (red line).}
\label{fig:dSCS_Cyl}
\end{figure}

Then, we investigate the convergence of the radiation pattern intensity $ \left| \mathbf{E}_S^\infty \left( \theta, \phi \right) \right|^2$ as a function of the number of considered MIMs. Specifically, we  plot in Fig. \ref{fig:dSCS_Cyl} the quantity $\left| \mathbf{E}_S^\infty \left( \theta, \phi \right) \right|^2$ as a function of the inclination angle $\theta$ in the $zx$  plane (a) and in the $yz$  plane (b), for different values of $h_{max}$. The cylinder is excited by an $x$-polarized plane wave of unit intensity, incoming from the negative $z$- axis. Also in this case, the cylinder is excited by the plane wave of Eq. \ref{eq:PW} with $\lambda=1m$.
We assumed a permittivity $\varepsilon_r = 5 + 0.1 i$, and we performed the calculation using Eq. \ref{eq:ModeExpansion} with $\tau = 10^{-2}$ and by the PMCHWT formulation. The green line corresponds to the MIM-VIE solution computed using $h_{max}=34$ modes. It is apparent that we get a very poor agreement with the SIE solution for both the investigated scattering planes. By increasing the number of modes to $h_{max}=107$ (blue line), we obtain convergence toward the SIE solution only for the $xz$ plane, as the other plane still shows lack of convergence. Finally, we obtain very good agreement for both the planes for $h_{max}=208$.


Similarly to what we have done for a sphere, we now calculate the break-even value of $n_{\varepsilon_r}$, in correspondence of which the MIM-VIE method becomes computationally advantageous compared to a direct VIE solution for the considered cylinder using $\tau = 10^{-2}$. In this case, the cumulative time for assembling both matrices $\bf K$ and $\bf R$ on $25$ processors is the same for both approaches, namely $ t_{\text{ASM}} = 4h \, 10m $. The total time needed for the calculation of the  sets of eigenvalues and corresponding MIMs is $t_{\text{EIG}} = 1h \, 30m $. For the direct VIE solution, the factorization time $t_{\text{FCT}} = 31m$. Therefore we get $n_{\varepsilon_r} \approx 3$. In conclusion, if we need to calculate the scattering from more than three cylinders having the geometry specified above but three different material compositions the MIM-VIE approach is computational advantageous compared to the direct VIE solver.

\section{Conclusions}
\label{sec:Conclusions}

We introduced a volume integral formulation for the calculation of a set of modes independent of the permittivity of the scatterer, which we have denoted as Material Independent Modes (MIMs). The solution of the scattering problem by a homogeneous object can be expanded in terms of MIMs, leading to a natural separation of the contributions of the permittivity and of the geometry. Specifically, the expansion coefficients are simple rational functions of the permittivity, whereas the MIMs depend solely on the geometry.

The calculation of the eigenvalues associated to the MIMs of a sphere is validated against the approach of Ref. \cite{Forestiere16}. We showed through numerical examples that for objects of linear dimension comparable to the incident wavelength only a moderate number of MIM is needed to accurately describe the scattering cross section for a wide range of permittivities and the radiation pattern. This fact has been shown by investigating the scattering from a sphere and a cylinder of size comparable to the incident wavelength and validating the results  exploiting the standard Mie theory and the PMCHWT surface integral formulation. 

In addition, we show that, exploiting the fact that the MIMs are independent of the permittivity of the scatterer, the representation of the solution in terms of them may be computationally advantageous compared to direct solution of the volume integral equation when the scattered field
has to be computed for many different values of the dielectric permittivity, given the size and shape of the dielectric body.

Eventually, it is also worth to point out that the application of the present method is limited to objects of size comparable to the wavelength due to its memory requirements and computational burden, unless suitable parallel sparsification techniques are used.


\begin{thebibliography}{10}


\bibitem{Schaubert84}
D.~Schaubert, D.~Wilton, and A.~Glisson, ``A tetrahedral modeling method for
  electromagnetic scattering by arbitrarily shaped inhomogeneous dielectric
  bodies,'' {\em IEEE Transactions on Antennas and Propagation}, vol.~32,
  pp.~77--85, Jan 1984.

\bibitem{draine1994discrete}
B.~T. Draine and P.~J. Flatau, ``Discrete-dipole approximation for scattering
  calculations,'' {\em JOSA A}, vol.~11, no.~4, pp.~1491--1499, 1994.

\bibitem{Sun09}
L.~E. Sun and W.~C. Chew, ``A novel formulation of the volume integral equation
  for electromagnetic scattering,'' {\em Waves in Random and Complex Media},
  vol.~19, no.~1, pp.~162--180, 2009.

\bibitem{DalNegro09}
L.~{Dal Negro}, G.~Miano, G.~Rubinacci, A.~Tamburrino, and S.~Ventre, ``A fast
  computation method for the analysis of an array of metallic nanoparticles,''
  {\em IEEE Transactions on Magnetics}, vol.~45, pp.~1618--1621, March 2009.

\bibitem{Markkanen12}
J.~Markkanen, P.~Yla-Oijala, and A.~Sihvola, ``Discretization of volume
  integral equation formulations for extremely anisotropic materials,'' {\em
  Antennas and Propagation, IEEE Transactions on}, vol.~60, no.~11,
  pp.~5195--5202, 2012.

\bibitem{harrington1996field}
R.~F. Harrington and J.~L. Harrington, {\em Field computation by moment
  methods}.
\newblock Oxford University Press, 1996.

\bibitem{taflove2005computational}
A.~Taflove and S.~C. Hagness, {\em Computational electrodynamics}.
\newblock Artech house, 2005.

\bibitem{Fuchs75}
Fuchs, R. ``Theory of the optical properties of ionic crystal cubes," Physical review B \textbf{11} (4) 1732 1975

\bibitem{Rojas86}
R. Rojas and F. Claro, ``Electromagnetic response of an array of particles: Normal-mode theory," Phys. Rev. B \textbf{34}, 3730 1986.


\bibitem{Bergman78}
D.~J. Bergman, ``The dielectric constant of a composite material—a problem in
  classical physics,'' {\em Physics Reports}, vol.~43, no.~9, pp.~377--407,
  1978.

\bibitem{Bergman80}
D.~J. Bergman and D.~Stroud, ``Theory of resonances in the electromagnetic
  scattering by macroscopic bodies,'' {\em Physical Review B}, vol.~22, no.~8,
  p.~3527, 1980.

\bibitem{Abajo02}
F J Garcia De Abajo and A Howie, ``Retarded field calculation of electron energy loss in inhomogeneous dielectrics.'' {\em Physical Review B} {\bf 65} 11 115418 2002

\bibitem{Fredkin2003}
D.~R. Fredkin and I.~D. Mayergoyz, ``Resonant behavior of dielectric objects
  (electrostatic resonances),'' {\em Phys. Rew. Letters}, vol.~91, 2003.

\bibitem{Mayergoyz05}
I.~Mayergoyz, D.~Fredkin, and Z.~Zhang, ``Electrostatic (plasmon) resonances in
  nanoparticles,'' {\em Phys. Rev. B}, vol.~72, p.~155412, 2005.

\bibitem{Markel10}
V.~A. Markel, ``Pole expansion of the lorenz-mie coefficients,'' {\em Journal
  of Nanophotonics}, vol.~4, no.~1, pp.~041555--041555, 2010.

\bibitem{Forestiere09}
C.~Forestiere, G.~Miano, G.~Rubinacci, and L.~Dal~Negro, ``Role of aperiodic
  order in the spectral, localization, and scaling properties of plasmon modes
  for the design of nanoparticle arrays,'' {\em Phys. Rev. B}, vol.~79,
  p.~085404, Feb 2009.

\bibitem{Markel:95}
V.~A. Markel, ``Antisymmetrical optical states,'' {\em J. Opt. Soc. Am. B},
  vol.~12, pp.~1783--1791, Oct 1995.

\bibitem{Forestiere16}
C.~Forestiere and G.~Miano, ``Material-independent modes for electromagnetic
  scattering,'' {\em Phys. Rev. B}, vol.~94, p.~201406, Nov 2016.

\bibitem{pascale2017spectral}
M.~Pascale, G.~Miano, and C.~Forestiere, ``Spectral theory of electromagnetic  scattering by a coated sphere,''  J. Opt. Soc. Am. B \textbf{34} 7 1524, 2017.

\bibitem{Bergman16}
A.~Farhi and D.~J. Bergman, ``Electromagnetic eigenstates and the field of an
  oscillating point electric dipole in a flat-slab composite structure,'' {\em
  Physical Review A}, vol.~93, no.~6, p.~063844, 2016.

\bibitem{forestiere2017electromagnetic}
C.~Forestiere and G.~Miano, ``On the nanoparticle resonances in the full-retarded regime
,'' {\em Journal of Optics},vol.~19, no.~6, 2017.

\bibitem{Tamburrino15}
A. Tamburrino, Z. Su, N. Lei, L. Udpa, S. Udpa,
 {\em The Monotonicity Imaging Method for PECT}, Studies in Applied Electromagnetics and Mechanics, 40, 159-166, 2015.

\bibitem{Tamburrino17}
Z Su, S Ventre, L Udpa, A Tamburrino,
 {\em Monotonicity Based Imaging Method for Time-Domain Eddy Current Problems}, submitted for publication, 2017

\bibitem{Tamburrino02}
Tamburrino, A., Rubinacci, G., {\em A new non-iterative inversion method for electrical resistance tomography}, (2002) Inverse Problems, 18 (6), pp. 1809-1829.


\bibitem{gastine1967electromagnetic}
M.~Gastine, L.~Courtois, and J.~L. Dormann, ``Electromagnetic resonances of
  free dielectric spheres,'' {\em IEEE Transactions on Microwave Theory and
  Techniques}, vol.~15, no.~12, pp.~694--700, 1967.

\bibitem{conwell1984resonant}
P.~R. Conwell, P.~W. Barber, and C.~K. Rushforth, ``Resonant spectra of
  dielectric spheres,'' {\em JOSA A}, vol.~1, no.~1, pp.~62--67, 1984.

\bibitem{Hill:88}
H.~S. C and B.~R. E, ``Morphology-dependent resonances,'' in {\em Optical
  Effects Associated with Small Particles} (P.~W. Barber and R.~K. Change,
  eds.), ch.~1, Singapore: World Scientific, 1988.

\bibitem{kristensen2013}
P.~T. Kristensen and S.~Hughes, ``Modes and mode volumes of leaky optical
  cavities and plasmonic nanoresonators,'' {\em ACS Photonics}, vol.~1, no.~1,
  pp.~2--10, 2013.

\bibitem{kristensen2015normalization}
P.~T. Kristensen, R.-C. Ge, and S.~Hughes, ``Normalization of quasinormal modes
  in leaky optical cavities and plasmonic resonators,'' {\em Physical Review
  A}, vol.~92, no.~5, p.~053810, 2015.

\bibitem{Garbacz65}
R.~J. Garbacz, ``Modal expansions for resonance scattering phenomena,'' {\em
  Proceedings of the IEEE}, vol.~53, pp.~856--864, Aug 1965.

\bibitem{garbacz1971generalized}
R.~Garbacz and R.~Turpin, ``A generalized expansion for radiated and scattered
  fields,'' {\em IEEE Transactions on Antennas and Propagation}, vol.~19,
  no.~3, pp.~348--358, 1971.

\bibitem{harrington1971theory}
R.~Harrington and J.~Mautz, ``Theory of characteristic modes for conducting
  bodies,'' {\em IEEE Transactions on Antennas and Propagation}, vol.~19,
  no.~5, pp.~622--628, 1971.

\bibitem{harrington1972characteristic}
R.~Harrington, J.~Mautz, and Y.~Chang, ``Characteristic modes for dielectric
  and magnetic bodies,'' {\em IEEE Transactions on Antennas and Propagation},
  vol.~20, no.~2, pp.~194--198, 1972.

\bibitem{chang1977surface}
Y.~Chang and R.~Harrington, ``A surface formulation for characteristic modes of
  material bodies,'' {\em IEEE Transactions on Antennas and Propagation},
  vol.~25, no.~6, pp.~789--795, 1977.

\bibitem{Chen15}
Y.~Chen and C.-F. Wang, {\em Characteristic Modes: Theory and Applications in
  Antenna Engineering}.
\newblock John Wiley and Sons, 2015.

\bibitem{bucci1995use}
O.~Bucci and G.~{Di Massa}, ``Use of characteristic modes in
  multiple-scattering problems,'' {\em Journal of Physics D: Applied Physics},
  vol.~28, no.~11, p.~2235, 1995.

\bibitem{grat2006}
G.~Rubinacci and A.~Tamburrino, ``A broadband volume integral formulation based
  on edge-elements for full-wave analysis of lossy interconnects,'' {\em IEEE
  Transactions on Antennas and Propagation}, vol.~54, pp.~2977--2989, Oct 2006.

\bibitem{ragr1988}
R.~Albanese and G.~Rubinacci, ``Solution of three dimensional eddy current
  problems by integral and differential methods,'' {\em IEEE Transactions on
  Magnetics}, vol.~24, pp.~98--101, Jan 1988.

\bibitem{ragr1998}
R.~Albanese and G.~Rubinacci, ``Finite element methods for the solution of 3d
  eddy current problems,'' vol.~102 of {\em Advances in Imaging and Electron
  Physics}, pp.~1 -- 86, Elsevier, 1997.

\bibitem{lehoucq1998arpack}
R.~B. Lehoucq, D.~C. Sorensen, and C.~Yang, {\em ARPACK users' guide: solution
  of large-scale eigenvalue problems with implicitly restarted Arnoldi
  methods}.
\newblock SIAM, 1998.

\bibitem{bohren08}
C.~F. Bohren and D.~R. Huffman, {\em Absorption and scattering of light by
  small particles}.
\newblock John Wiley \& Sons, 2008.

\bibitem{WriedtBook}
A.~Doicu, T.~Wriedt, and Y.~Eremin, {\em Light Scattering by Systems of
  Particles}.
\newblock Springer-Verlag, 2006.

\bibitem{Bohren1998}
C.~F. Bohren and D.~R. Huffman, {\em Absorption and Scattering of Light by
  Small Particles}.
\newblock Wiley, 1998.

\bibitem{blackford1997scalapack}
L.~S. Blackford, J.~Choi, A.~Cleary, E.~D'Azevedo, J.~Demmel, I.~Dhillon,
  J.~Dongarra, S.~Hammarling, G.~Henry, A.~Petitet, {\em et~al.}, {\em
  ScaLAPACK users' guide}.
\newblock SIAM, 1997.





\end{thebibliography}
\end{document}